\documentclass{article}
\usepackage{spconf}
\PassOptionsToPackage{numbers,sort&compress}{natbib}
\usepackage{amsmath, amssymb, amsfonts}
\usepackage{algorithmic}
\usepackage{graphicx}
\usepackage{makecell}
\usepackage{hyperref}
\usepackage{textcomp}
\usepackage{xcolor}
\usepackage{booktabs}
\usepackage{multirow}
\usepackage{array}

\usepackage{adjustbox}
\usepackage{kotex}
\usepackage{subcaption}
\usepackage{caption}
\usepackage[utf8]{inputenc}

\ninept
\begin{document}
\title{Acoustic-to-Text KV Compression for Full-Duplex Speech Models
}

\name{Yejin Lee, Seungbeom Kim, Yongha Lee, Kyuhong Shim}
\address{ Sungkyunkwan University, Republic of Korea \\
    \texttt{\small yj.lee@skku.edu, khshim@skku.edu} }

\maketitle
\begin{abstract}
    Full-duplex speech language models continuously accumulate acoustic key--value (KV) states, making long-running interactions memory-intensive.
During listening, the model can finish processing an audio unit before the next arrives; we term the remaining interval listening-time slack.
We propose acoustic-to-text KV compression, which introduces a transcription side channel to convert incoming speech into compact textual memory within this interval. 
When the cache exceeds a target budget during inference, older acoustic states are evicted while transcripts and recent acoustic context remain.
We train the side channel with LoRA using cross-entropy on transcription segments.
To preserve listening and speaking behavior, we apply knowledge distillation to the original model’s token-level output distributions at native prediction positions.
On ten-minute LongSpeech sessions, our MiniCPM-o 4.5 implementation reduces peak streaming KV-cache size by 64.6\% compared with the same model without eviction. 
The proposed method also improves transcription, temporal question answering, and summarization over the baseline.
Full-Duplex-Bench evaluations further show comparable pause-handling, turn-taking, and interruption performance.
\end{abstract}


\section{Introduction}\label{sec:intro}

\begin{figure*}[t!]
    \centering
    \includegraphics[width=0.95\linewidth]{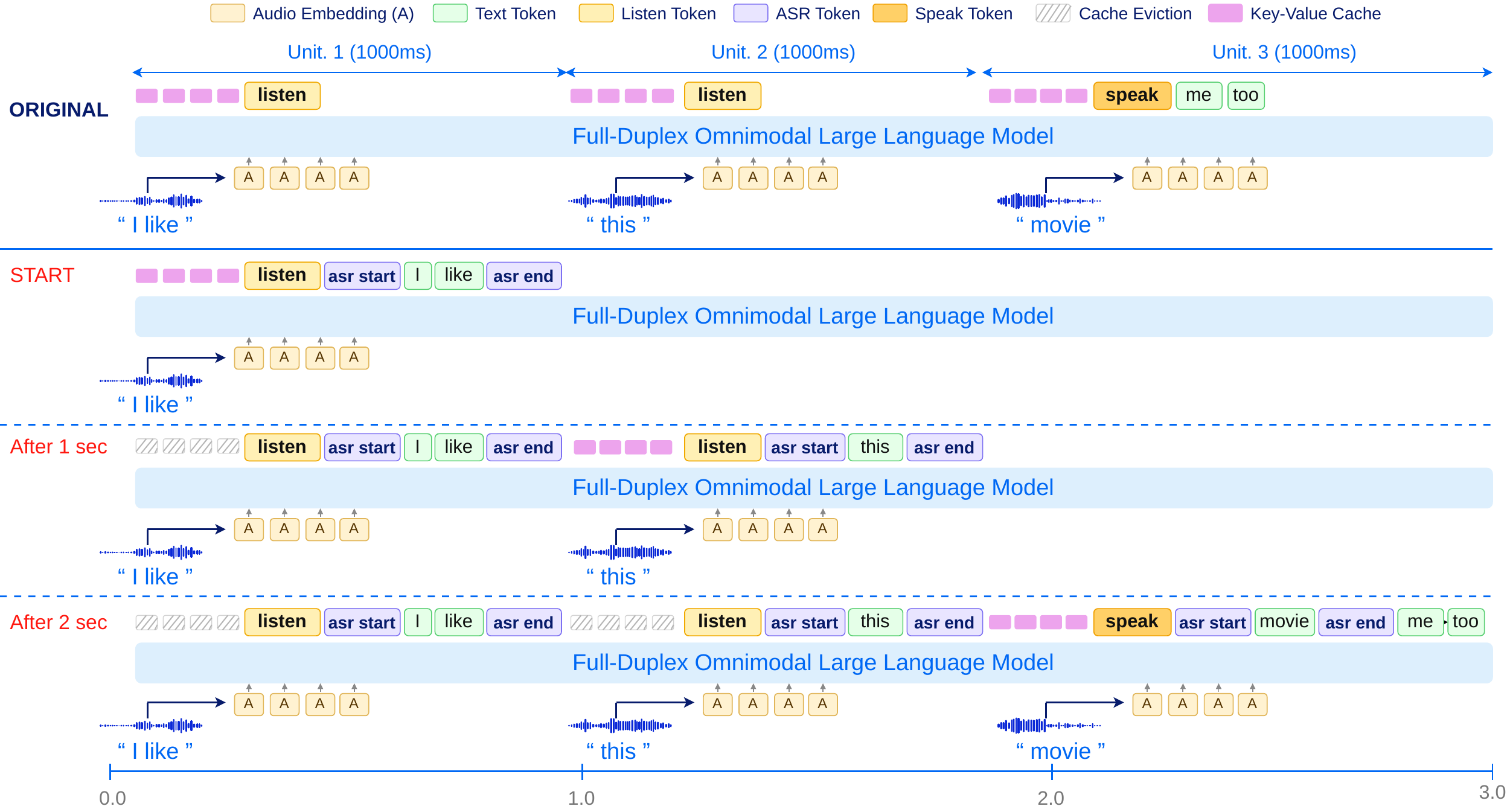}
    \caption{
    Overview of acoustic-to-text KV compression: native streaming (top) and the proposed method (lower rows).
    The added transcription side channel uses listening-time slack to generate transcript tokens that remain in context but are excluded from spoken output.
    When the KV cache exceeds a target budget, older acoustic states are evicted while transcripts and recent acoustic context are retained.
    }
    \label{fig:concept}
\end{figure*}

Full-duplex speech language models process incoming speech while generating responses, supporting continuous spoken interaction~\cite{synchronousllm,fullduplexschema,moshi,omniflatten}. 
As interaction progresses, earlier speech provides context for subsequent responses, and the relevance of past content may become apparent later.
However, retaining this history in acoustic form continuously expands the key–value (KV) cache because speech representations occupy substantially more sequence positions than text~\cite{speechxl,lu2026speechkv}.
Compression applied only after input completion cannot reduce the peak memory already incurred during input processing.
Furthermore, long-running spoken interactions require control of peak KV cache usage throughout streaming~\cite{kim2024infinipot}.
Such memory control must retain information from past speech while preserving real-time listening and speaking behavior.

To retain past speech in a compact form while meeting real-time deadlines, we exploit the computation time available between incoming audio units. 
MiniCPM-o 4.5~\cite{cui2026minicpm} processes one-second audio units and decides whether to continue listening or start speaking after each unit.
During listening, processing can finish before the next unit arrives; we term the remaining interval \textit{listening-time slack}.
In our setup, this interval is approximately 900 ms, providing sufficient time for incremental automatic speech recognition (ASR). 
The resulting transcript occupies approximately three sequence positions per second of speech, compared with 17 for the acoustic history.

We propose \textit{acoustic-to-text KV compression}, which introduces a transcription side channel to record incoming speech during listening-time slack.
Special ASR tokens delimit each transcription segment, which remains in the language-model context and is excluded from spoken output.
When the resident KV cache exceeds a target budget, we evict older acoustic states while retaining transcript tokens. 
We maintain a recent acoustic window to accommodate transcription delay before eviction. 
The retained transcript stores the linguistic content of earlier speech, while recent acoustic context remains available for ongoing interaction.
The side channel reuses the existing language-model backbone and requires no separate ASR model at inference.

Adapting the model for transcription must also preserve native full-duplex listening and speaking behavior.
We train the side channel with low-rank adaptation (LoRA) and word-level forced-alignment supervision.
However, transcription-only training can disrupt native conversational behavior.
We therefore use the original model as a frozen teacher that processes the same audio without transcription segments. 
Cross-entropy supervises transcription segments, while token-level knowledge distillation aligns the student’s output distributions with the teacher’s at corresponding native prediction positions.
This adaptation uses only short transcribed utterances, avoiding the need for long-form speech training or full-duplex interaction data collection.

Experiments on MiniCPM-o 4.5 evaluate peak KV cache usage, long-form speech understanding, and real-time interaction. 
On ten-minute LongSpeech sessions~\cite{yang2026longspeech}, peak streaming KV cache size decreases by 64.6\% compared with the same adapted model without eviction.
The proposed method also improves transcription, temporal question answering, and summarization performance over native streaming. 
At a 4K KV cache budget, processing, including transcription and cache eviction, averages 461 ms per one-second audio unit on an H100 GPU with batch size 1.
No deadline misses are observed among the 10,039 measured units.
Full-Duplex-Bench~\cite{lin2025fdb} evaluations further show comparable pause-handling, turn-taking, and interruption performance.

\noindent Our contributions are threefold:
\begin{itemize}
    \item We propose acoustic-to-text KV compression for full-duplex speech language models. A newly introduced transcription side channel uses listening-time slack to construct textual memory before acoustic states are evicted during streaming.
    \item We combine transcription supervision with token-level knowledge distillation at native prediction positions to train the side channel with LoRA while constraining changes to listening and speaking behavior.
    \item We demonstrate lower peak KV cache usage and improved long-form speech understanding, with real-time processing and comparable pause-handling, turn-taking, and interruption performance.
\end{itemize}

\section{Related Work}
\label{sec:related}

\subsection{Full-Duplex Speech Models}

Full-duplex speech models process incoming speech while generating responses, supporting continuous spoken interaction.
SyncLLM~\cite{synchronousllm} predicts time-synchronized chunks of user and agent speech in one decoder, LSLM~\cite{ma2025language} adds a streaming listening channel to a speaking model so that it can be interrupted, and Wang et al.~\cite{fullduplexschema} switch between speaking and listening with control tokens predicted by the language model.
Moshi~\cite{moshi} models user and agent speech as parallel streams, OmniFlatten~\cite{omniflatten} adapts a text language model to full-duplex interaction, and MiniCPM-o~4.5~\cite{cui2026minicpm} aligns multimodal inputs and outputs along a shared temporal axis and decides between listening and speaking once per second.
Benchmarks such as Full-Duplex-Bench~\cite{lin2025fdb} evaluate turn-taking, backchannel, and interruption behavior in full-duplex speech models.

\subsection{KV Cache Compression}

KV cache compression has been widely studied in text-based large language models to reduce the memory cost of long-context inference.
StreamingLLM~\cite{streamingllm} keeps attention-sink tokens and a recent window, while H$_2$O~\cite{zhang2023h2o} and SnapKV~\cite{li2024snapkv} select states using attention scores.
InfiniPot~\cite{kim2024infinipot} repeatedly compresses the cache during input processing to operate within a fixed memory budget.

Compression for speech language models has also attracted growing interest.
FastAdaSP~\cite{lu2024fastadasp} reduces the number of speech tokens, while Lu et al.~\cite{lu2026speechkv} compress the KV cache instead of the speech embeddings.
AudioKV~\cite{wang2026audiokv} allocates the acoustic KV budget across attention heads.
Speech-XL~\cite{speechxl} processes speech in intervals and retains learned summary-token KV states after discarding the corresponding acoustic states.
These speech compression methods retain compressed speech representations rather than explicit transcripts.
Our method constructs textual memory during listening and evicts older acoustic states during full-duplex streaming.

\subsection{Computation During Listening}

Recent speech language models, including full-duplex ones~\cite{wu2026chronological,wu2026the}, also compute while the speech input is still arriving, instead of waiting for it to end.
Shih et al.~\cite{think_while_listening} initiate text-space reasoning from partial speech, while SHANKS~\cite{chiang2026shanks} generates unspoken chain-of-thought reasoning after each incoming speech chunk.
Song et al.~\cite{song2026learning} train a wait–think–answer controller to schedule intermediate reasoning from partial audio, whereas FLAIR~\cite{wu2026the} maintains a recurrent latent reasoning state during speech perception in a full-duplex model.
These methods use computation during listening to improve reasoning and response quality.
Our method adds a transcription side channel to an existing full-duplex model through lightweight LoRA fine-tuning.
The side channel uses listening-time slack to construct textual memory that remains available after acoustic KV eviction.

\begin{table*}[tb]
\centering
\caption{LongSpeech results on 100 ten-minute sessions per task. Listening and Query denote resident KV positions after streaming and during temporal QA.
$^{\dagger}$Native streaming. $^{\ddagger}$StreamingLLM (sink + 2,000 recent positions). $^{\P}$External ASR (offline Whisper on 30-s chunks, or the NeMo streaming FastConformer with a 1.04-s lookahead).
}
\label{tab:main_table1}
\resizebox{1.0\linewidth}{!}{
\setlength\tabcolsep{11pt}
\renewcommand{\arraystretch}{1.0}
\begin{tabular}{@{}l c c c cc cc@{}}
\toprule
& \multicolumn{1}{c}{\textbf{Added params}}
& \multicolumn{1}{c}{\textbf{ASR}}
& \multicolumn{1}{c}{\textbf{Temporal QA}}
& \multicolumn{2}{c}{\textbf{Summary}}
& \multicolumn{2}{c}{\textbf{KV Cache}}  \\
\cmidrule(lr){2-2}
\cmidrule(lr){3-3}
\cmidrule(lr){4-4}
\cmidrule(lr){5-6}
\cmidrule(l){7-8}

Method
& 
& WER $\downarrow$
& Accuracy $\uparrow$  % GPT-4o judge
& BLEU $\uparrow$
& R-1 $\uparrow$
& Listening $\downarrow$ 
& Query $\downarrow$ \\
\midrule

MiniCPM-o 4.5 $^{\dagger}$
& --
& 96.9 & 30.0 & 2.7 & 27.3 & 12.3K & 10.4K \\

MiniCPM-o 4.5 $^{\ddagger}$
& --
& 109.5 & 4.0 & 2.1 & 28.3 & 4.5K & 2.0K \\
\midrule

Whisper-tiny cascade $^{\P}$
& 39M
& 15.9 & 45.0 & 10.5 & 48.9 & 12.3K & 2.1K \\

FastConformer streaming cascade $^{\P}$
& 115M
& 10.3 & 40.0 & \textbf{11.6} & 49.5 & 12.3K & 1.9K \\
Whisper-large-v3 cascade $^{\P}$
& 1.55B
& \textbf{9.35} & \textbf{47.0} & 10.8 & \textbf{49.6} & 12.3K & 2.1K \\
\midrule
\textbf{Ours}
& 43.7M
& 12.9 & 42.0 & 11.0 & 49.0 & \textbf{4.0K} & \textbf{1.8K} \\

\bottomrule
\end{tabular}%
}
\end{table*}

\begin{table*}[tb]
\centering
\caption{
Full-Duplex-Bench v1.0 results under identical decoding settings. TOR denotes takeover rate, including the pause columns. Ours w/o distillation uses transcription cross-entropy alone.
}
\label{tab:fdb}

\resizebox{1.0\linewidth}{!}{
\setlength\tabcolsep{10pt}
\renewcommand{\arraystretch}{1.0}
\begin{tabular}{@{}l c cc cc ccc@{}}
\toprule
& \multicolumn{1}{c}{\textbf{Backchannel}}
& \multicolumn{2}{c}{\textbf{Pause}}
& \multicolumn{2}{c}{\textbf{Turn-taking}}
& \multicolumn{3}{c}{\textbf{Interruption}} \\

\cmidrule(lr){2-2}
\cmidrule(lr){3-4}
\cmidrule(lr){5-6}
\cmidrule(l){7-9}

Model
& TOR $\downarrow$
& Candor $\downarrow$
& Synthetic $\downarrow$
& TOR $\uparrow$
& Latency (s) $\downarrow$
& Score $\uparrow$
& TOR $\uparrow$
& Latency (s) $\downarrow$ \\

\midrule
MiniCPM-o 4.5 & 0.000 & 0.088 & 0.146 & 0.664 & 1.944 & 4.65 & 0.975 & 1.795 \\
\midrule
\textbf{Ours} w/o distillation & 0.982 & 0.981 & 0.978 & 1.000 & 0.000 & 0.19 & 0.946 & 0.452 \\
\textbf{Ours}          & 0.218 & 0.102 & 0.117 & 0.706 & 1.965 & 4.54 & 0.940 & 1.820 \\

\bottomrule
\end{tabular}%
}
\end{table*}

\begin{table*}[tb]
\centering
\caption{
KV-budget ablation of the same adapted model. End and Peak report final and peak streaming KV positions. Unbounded disables eviction; 0 (floor) evicts all eligible states. Unit time is the wall-clock time per one-second unit including side-channel decoding and eviction (one H100 80GB, batch size 1, 15 sessions per budget); $>$1 s reports deadline misses. Bold marks the selected 4K budget.
}
\label{tab:kv_budget_table3}

\resizebox{1.0\linewidth}{!}{
\setlength\tabcolsep{9pt}
\renewcommand{\arraystretch}{1.0}
\begin{tabular}{@{}l cc c c ccc ccc@{}}
\toprule
& \multicolumn{2}{c}{\textbf{KV Cache}}
& \multicolumn{1}{c}{\textbf{ASR}}
& \multicolumn{1}{c}{\textbf{Temporal QA}}
& \multicolumn{3}{c}{\textbf{Summary}}
& \multicolumn{3}{c}{\textbf{Unit time (ms)}} \\

\cmidrule(lr){2-3}
\cmidrule(lr){4-4}
\cmidrule(lr){5-5}
\cmidrule(lr){6-8}
\cmidrule(l){9-11}

Ours (budget)
& End $\downarrow$
& Peak $\downarrow$
& WER $\downarrow$
& Accuracy $\uparrow$ % GPT-4o Judge
& BLEU $\uparrow$
& R-1 $\uparrow$
& R-L $\uparrow$
& Mean $\downarrow$
& p99 $\downarrow$
& $>$1 s $\downarrow$ \\

\midrule
Unbounded
& 11.35K & 11.35K
& 14.4 & 42.0 & 11.73 & 48.5 & 28.6 & 458 & 683 & 0.32\% \\

8K
& 7.99K & 8.02K
& 13.6 & 45.0 & 11.18 & 49.2 & 28.5 & 440 & 636 & 0.03\% \\

6K
& 5.99K & 6.02K
& 12.9 & 46.0 & 11.53 & 49.2 & 28.7 & 451 & 669 & 0.05\% \\

\textbf{4K}
& \textbf{3.99K}
& \textbf{4.02K}
& \textbf{12.9}
& \textbf{42.0}
& \textbf{11.01}
& \textbf{49.0}
& \textbf{28.1}
& \textbf{461} & \textbf{675} & \textbf{0.00\%} \\

0 (floor)
& 1.92K & 1.94K
& 19.7 & 41.0 & 11.28 & 49.2 & 28.6 & 501 & 1337 & 4.07\% \\

\bottomrule
\end{tabular}%
}
\vspace{-0.1cm}
\end{table*}

\section{Acoustic-to-Text KV Compression}
\label{sec:method}

\subsection{Transcription Side Channel}
\label{sec:sidechannel}

We introduce a transcription side channel that represents incoming speech as textual memory for subsequent acoustic KV eviction.
MiniCPM-o~4.5 processes speech in duplex units of duration $\Delta$=1 second, each containing a \texttt{<unit>} marker, approximately ten audio tokens, and a native \texttt{<|listen|>} or \texttt{<|speak|>} decision that either closes the unit or starts a response.
We refer to the corresponding decoding paths as LISTEN and SPEAK units.

For each unit $k$, we place a transcription segment $y_k$, delimited by the newly introduced \texttt{<|asr\_start|>} and \texttt{<|asr\_end|>} tokens, immediately after the first native decision.
In both LISTEN and SPEAK units, transcription uses the listening-time slack before the next audio unit arrives.
In SPEAK units, native response generation continues after the transcription segment.
The segment remains in the language-model context but is excluded from the speech decoder input and the spoken response.

At inference, the model predicts whether to open the side channel.
If the \texttt{<|asr\_start|>} token is predicted, we decode greedily up to 64 text tokens until \texttt{<|asr\_end|>}.
If the limit is reached or a control token appears, we close the segment with a forced \texttt{<|asr\_end|>}.
If \texttt{<|asr\_start|>} is not predicted, the unit contains no transcription and native decoding continues.
Errors caused by skipped or truncated segments are included in the reported WER.

\subsection{Transcription Supervision and Behavior Distillation}
\label{sec:distill}

We supervise transcription with cross-entropy and constrain changes to native listening and speaking behavior through token-level knowledge distillation from the frozen original model (i.e., the teacher).
We first construct transcription targets from word-level forced alignments.
For a reference word $w_i$ with acoustic endpoint $e_i$, we assign the word to unit $k_i = \left\lceil \frac{e_i+\delta}{\Delta} \right\rceil$, where $\delta$ is an additional emission delay.
Each word appears in exactly one target segment.

For each training utterance, the teacher processes the same audio without transcription segments.
We collect its full-vocabulary output distributions at native prediction positions, extending them to the student vocabulary by assigning zero probability to the two added ASR tokens.
The student is teacher-forced on the native token sequence, with the supervised transcription segments inserted after the native decisions.
Teacher and student predictions are aligned by native prediction events rather than sequence indices, since the inserted segments shift the student’s prediction positions.

The training objective is a sum of four loss terms:
\begin{equation}
\mathcal{L}
= \mathcal{L}_{\mathrm{ASR}}
+ \mathcal{L}_{\mathrm{native}}
+ \mathcal{L}_{\mathrm{dec}}
+ \mathcal{L}_{\mathrm{turn}},
\end{equation}
where $\mathcal{L}_{\mathrm{ASR}}$ is masked cross-entropy over the transcription tokens and their delimiters.
At each aligned native prediction position, we compute the forward divergence
$T^2 D_{\mathrm{KL}}\left(P_T^{(T)} \,\Vert\, P_S^{(T)}\right)$,
where $P_T^{(T)}$ and $P_S^{(T)}$ denote the teacher and student distributions at temperature $T$=1, respectively.
The loss $\mathcal{L}_{\mathrm{native}}$ averages this divergence over all valid native prediction positions.

Averaging over all native prediction positions can underweight infrequent decision types and turn boundaries. We therefore introduce group-balanced auxiliary losses that reweight the same per-position KL divergences to emphasize these events.
For $\mathcal{L}_{\mathrm{dec}}$, the first decision in each unit is grouped as listen, speak, or other according to the token actually fed.
For $\mathcal{L}_{\mathrm{turn}}$, non-decision positions are grouped as turn-end or turn-continuation according to whether the fed token is \texttt{<|turn\_eos|>}.
Each auxiliary loss first averages within each group and then averages across the groups present.

We train rank-16 LoRA adapters in the language model together with the embedding and output rows of the two new ASR tokens; the audio encoder, vision modules, speech decoder, and teacher stay frozen.
Adaptation uses only short transcribed utterances, without requiring long-form speech or full-duplex interaction training data.

\subsection{Online KV Eviction}
\label{sec:consolidation}

After each duplex unit, we compare the number of resident KV positions, including the system prefix and text entries, with a budget $B$ (4K in our main experiments).
When the cache exceeds $B$, we evict acoustic and structural KV states from the oldest unit outside the five most recent units, called the \textit{retention window}.
Eviction repeats until the cache fits within $B$ or no eligible units remain.
The protected window accommodates transcription delay: the duration of five seconds covers the measured emission latency of the side channel on LibriSpeech test-clean.

Eviction removes the KV states associated with the unit marker, audio tokens, native decision, ASR delimiters, and terminators, while retaining transcript tokens, model-generated response text, and the system prefix.
Position IDs advance with every token fed to the model, regardless of eviction, so the remaining keys keep their original rotary positions without re-encoding.

\section{Experimental Results}
\label{sec:experiments}

\subsection{Training}
We fine-tune the model on 132K short utterances from the 460-hour LibriSpeech~\cite{panayotov2015librispeech} \texttt{train-clean} subset, using word alignments from the Montreal Forced Aligner (MFA) with an emission delay of $\delta=0.5$\,s.
LoRA (rank 16, $\alpha=32$) on the attention and feed-forward projections adds 43.7M parameters (0.53\% of the language model).
Training runs for three epochs with a learning rate of $10^{-4}$ and a batch size of 16, with unit weight on all four loss terms.

\subsection{Evaluation}
We evaluate long-running inference on LongSpeech~\cite{yang2026longspeech}, a benchmark of ten-minute audio sessions with three tasks: long-form transcription (WER), temporal question answering about the order of events (accuracy), and summarization (BLEU and ROUGE).
We use the first 100 English sessions of each task, with QA accuracy judged by GPT-4o.
External-ASR cascades answer from re-encoded transcripts while retaining the native acoustic cache during streaming. 
For our method, task queries use only the incoming-speech transcript, re-encoded after discarding the streaming cache.
We also measure conversational behavior on Full-Duplex-Bench~\cite{lin2025fdb} v1.0 using the official protocol and scoring scripts on all 727 samples of its five tasks.

\subsection{Long-form Speech Understanding}

Table~\ref{tab:main_table1} shows lower WER (12.9 versus 96.9) and higher temporal QA accuracy (42\% versus 30\%) than native streaming, together with improved summarization. 
In particular, the StreamingLLM comparison supports preserving an explicit linguistic record before eviction, allowing subsequent queries to access content beyond the recent acoustic window. 
Our method achieves comparable QA and summarization performance to external-ASR cascades, although FastConformer and Whisper-large-v3 achieve lower WER with more added parameters.
Constructing this record within the duplex backbone integrates transcription with streaming cache management without requiring a separate ASR model.
The results demonstrate that adaptation on short utterances supports ten-minute speech processing without long-form training.

\subsection{Full-Duplex Interaction}

Table~\ref{tab:fdb} shows comparable pause-handling, turn-taking, and interruption performance, with a higher backchannel takeover rate (0.218 versus 0.000). Without distillation, the model takes over nearly all pauses and backchannels and starts speaking from the first unit in almost every sample. The resulting turn-taking TOR of 1.0 and zero latency reflect premature speech. Distillation constrains this tendency and helps retain native conversational behavior.

The ablation shows that updating only a small fraction of parameters can still disrupt the pretrained interaction policy, motivating explicit behavioral constraints during transcription adaptation. 
The frozen teacher supplies this supervision on ordinary transcribed utterances, allowing the side channel to be trained without additional full-duplex interaction data.

\subsection{Streaming Memory and Latency}

At a 4K budget, peak streaming KV size decreases from 11.35K to 4.02K positions, a 64.6\% reduction relative to the same adapted model without eviction (Table~\ref{tab:kv_budget_table3}).
QA and summarization scores remain close to the unbounded setting, while WER improves from 14.4 to 12.9.
More aggressive eviction at the floor raises WER to 19.7, supporting 4K as the operating point.

On 2,620 LibriSpeech test-clean utterances, mean word-emission delay is 1.01 s, compared with 9.5 s for native prompted transcription. 
The measured maximum of 2.59 s for our method fits within the five-second retention window. 
At 4K, unit processing averages 461 ms, with a 99th percentile of 675 ms and no deadline misses across 10,039 one-second units. 
This operating point controls streaming KV memory without a separate ASR model while accommodating transcription and eviction within the real-time processing budget.

\section{Conclusion}
We proposed acoustic-to-text KV compression for full-duplex speech language models. 
We introduced a transcription side channel that uses listening-time slack to construct textual memory before online acoustic eviction, without a separate ASR model.
We combined transcription supervision with token-level distillation to train LoRA adapters on short transcribed utterances while constraining changes to native listening and speaking behavior. 
Experiments demonstrated a 64.6\% reduction in peak streaming KV usage on MiniCPM-o 4.5 relative to the same adapted model without eviction, alongside improved long-form speech understanding over native streaming.
We further showed real-time processing and comparable pause-handling, turn-taking, and interruption performance through timing and Full-Duplex-Bench evaluations.

\newpage
\section{COMPLIANCE WITH ETHICAL STANDARDS}
This study used existing publicly available datasets and models, and
no new data were collected. No ethical approval was required.
\bibliographystyle{IEEEbib}
\bibliography{reference}
\end{document}